\documentclass{svproc}

\usepackage{url}

\usepackage{hyperref}
\usepackage{graphicx}
\usepackage[utf8]{inputenc}
\usepackage{soul,color}
\graphicspath{{Figures/}}  

\usepackage[font=small,labelfont=bf,labelsep=period]{caption} 
\renewcommand{\figurename}{\textbf{Fig.}} 
\usepackage{subcaption} 
\usepackage{amsmath}
\usepackage{wrapfig}
\usepackage[scr=rsfs,cal=boondox]{mathalfa}  
\usepackage{bm} 
\usepackage{tensor} 

\usepackage{array}
\newcolumntype{L}[1]{>{\raggedright\let\newline\\\arraybackslash\hspace{0pt}}m{#1}}
\newcolumntype{C}[1]{>{\centering\let\newline\\\arraybackslash\hspace{0pt}}m{#1}}
\newcolumntype{R}[1]{>{\raggedleft\let\newline\\\arraybackslash\hspace{0pt}}m{#1}}

\newcommand{\secreff}[1]{\hyperref[#1]{Section \ref*{#1}}} 
\newcommand{\figreff}[1]{\hyperref[#1]{Fig. \ref*{#1}}} 
\newcommand{\tabreff}[1]{\hyperref[#1]{Table. \ref*{#1}}} 

\begin{document}
\mainmatter              
\title{Monitoring the Seismic Behavior of a Scaled RC Frame with Intermediate Ductility in a Shaking Table Test}
\titlerunning{ }  
%
\author{Mohammad Vasef\inst{1,2} \and Mohammad Sadegh Marefat\inst{2} \and \\ Seyed Sina Shid-Moosavi\inst{1} \and Peng ``Patrick" Sun\inst{1,*}}
\authorrunning{ } 
%
\tocauthor{ }
%

\institute{Department of Civil, Environmental, and Construction Engineering, \\ University of Central Florida, Orlando FL 32816, USA,\\ 
\and
School of Civil Engineering, College of Engineering, University of Tehran, \\ P.O. Box 11365-4563, Tehran, Iran. \\
\email{*peng.sun@ucf.edu}}

\maketitle              

\begin{abstract}
One of the commonly used seismic force-resisting systems in structures is Reinforced Concrete (RC) Intermediate  Moment Frames (IMF).  Although using the IMF is not allowed in high seismic hazard zones according to ASCE 7-10, it is permitted in both Iran’s 2800 Seismic Standard and New Zealand’s Seismic Code.  This study investigates the seismic behavior of a reinforced concrete IMF subjected to earthquake excitations using shaking table tests on a 2D RC structural model which is designed under the regulations of ACI318-19.  The scale factor of 1/2.78 is selected for the frame fabrication due to the size limit of the shaking table. The constructed model has three stories with a height as 115 cm for each story, the clear length of beams as 151 cm, and cross-sectional dimensions of columns and beams as 11×11 cm and 12×11 cm, respectively.  The whole structure is supported by a foundation that is 173 cm long, 52 cm wide, and 22 cm deep.  Columns and beams are reinforced with 8 mm diameter longitudinal ribbed bars and stirrups with 6 mm diameter. The tests are conducted in stages with increasing peak ground acceleration (PGA) till the failure of the frame. Sarpol-E-Zahab earthquake seismic record is adopted for the experiment.  The structural responses (\textit{e.g.,} displacements, longitudinal bars’ strain, crack propagation, accelerations) are monitored during the test using both conventional sensors and vison-based sensors. As a comparative study, both conventional sensors and computer vision techniques are used to monitor the health state and to analyze the structural dynamics of the scaled RC frame structure.
\keywords{shaking table test, structural health monitoring, computer vision, intermediate moment frame}
\end{abstract}
\section{Introduction}
Moment-resisting frames are one of the most commonly used structural systems all around the world which provide resistance to lateral forces (\textit{e.g.,} seismic load). Previous researchers have shown that RC structures' seismic behavior is influenced by many factors (\textit{e.g.,} sections' geometry, material properties, and connection rigidity) ~\cite{shiravand2021effect,shid2018performance}. The resistance to lateral forces in IMF is provided mainly by the rigid connection of beams and columns and also by the flexural behavior of the members. There are three types of moment frames: Ordinary Moment Frames (OMF); Intermediate Moment Frames (IMF), and Special Moment Frames (SMF). According to Iran’s Standard No.2800-4th Edition (Iranian Code of Practice for Seismic Resistant Design of Buildings) and the New Zealand’s seismic code (NZS1170), using IMFs is permitted for ordinary buildings (with importance factors of 0.8, 1.0, and 1.2) in high seismic hazard zones (A=0.35g). While ASCE 7-10 do not permit to use these systems in high seismic zones. 
A large number of experimental studies on RC frames have been conducted in the past many years. Skjaerbaek et.al~\cite{skjaerbaek1997shaking} tested a series of shaking table experiments on seven 1/5 scaled RC frames to investigate system identification methods on time-varying systems and to compare different methods for damage assessment of RC structures. Quintana-Gallo et al.~\cite{quintana2010shake} investigated the feasibility of the previously developed RC frames and latterly improved retrofit solutions by a series of shake table tests on under-designed 1/2.5 scaled RC frames in different scenarios (\textit{e.g.,} with or without seismic retrofit intervention, with or without infill panels). Bayhan et al.~\cite{bayhan2012experimental} assessed the seismic behavior of RC frames with weak beam-column joints. The researchers used a two-story by two-bay 1/2.25 scaled RC frame which was subjected to earthquake excitation on a shaking table. Benavent-Climent et al.~\cite{benavent2014seismic} experimentally investigated a 2/5 scaled RC frame with hysteretic dampers using shake-table test. Benavent-Climent et al.~\cite{benavent2014seismic} concluded that combining hysteretic dampers with flexible reinforced concrete frames improves the seismic performance of conventional RC frames. Benavent-Climent et al.~\cite{benavent2014shake} conducted shake-table tests on a 2/5 scaled RC frame representing a conventional construction designed under current building code provisions in the Mediterranean area. A sequence of dynamic tests, including free vibrations and four seismic simulations, were used in the test. Gong et al.~\cite{gong2019comparing} compared the seismic behavior of two types of RC frame structures by applying shaking table tests: (1) an infilled RC frame without infill walls in the first story and (2) a bare RC frame.  The test models were two three-story, two-bay, 1/4 scaled RC frame structures.
The work in this paper presents an experimental investigation of a 2D, 1/2.78 scaled model of a three-story, one-bay concrete intermediate moment frame structure. It is designed under the regulations of ACI318-19 and tested on the shaking table in the Soil Laboratory at the University of Tehran. The seismic behavior under different levels of earthquake as well as the dynamic characteristics of the structure are investigated at the University of Central Florida.

%
\section{Material and Method}
\subsection{Design and Fabrication of RC Frame}
Since the cross-sections of members are in small scale, a proper concrete pouring and vibrating is important to avoid fabrication issues (\textit{e.g.,} air voids on the surface of concrete members). Because it is not feasible to fabricate the frame on the shake table directly, it is crucial to have a platform for transferring the RC frame to the table after curing. Hence, a steel frame is constructed and the RC frame can be placed on the steel frame for transporting. The steel frame makes the process of pouring, vibrating, and transferring much easier. Then, the formwork and reinforcement are installed and fixed. Finally, the concrete is poured into prepared PVC formwork and cured properly.\par
The three-story, one-bay, RC moment frame (Fig.1) is considered as the prototype structure in this study. The reinforcement design is based on the regulations of ACI318-19 and lateral seismic loading is chosen according to the Iran Standard NO.2800. The prototype building is assumed that: (i) it is located in the city of Tehran (Iran) where the design ground acceleration is A=0.35g (g is the gravity acceleration); (ii) its usage is residential; and (iii) it is located on soil (type II). The concrete compressive strength $f_{c}^{'}$ is assumed to be 25MPa and the yield strength of longitudinal bars is $f_y$=400MPa in the calculation process of the structural design. The cross-sections of the RC columns and beams are 30×30 $cm^2$ and 33×30 $cm^2$, respectively. The floor system consists of one-way joists as shown in Fig.1(a). According to the Iranian seismic code, the response modification coefficient (R factor) is set as 5. 
\begin{figure}[h]
     \centering
     \begin{subfigure}[b]{0.40\textwidth}
         \centering
         \includegraphics[width=\textwidth]{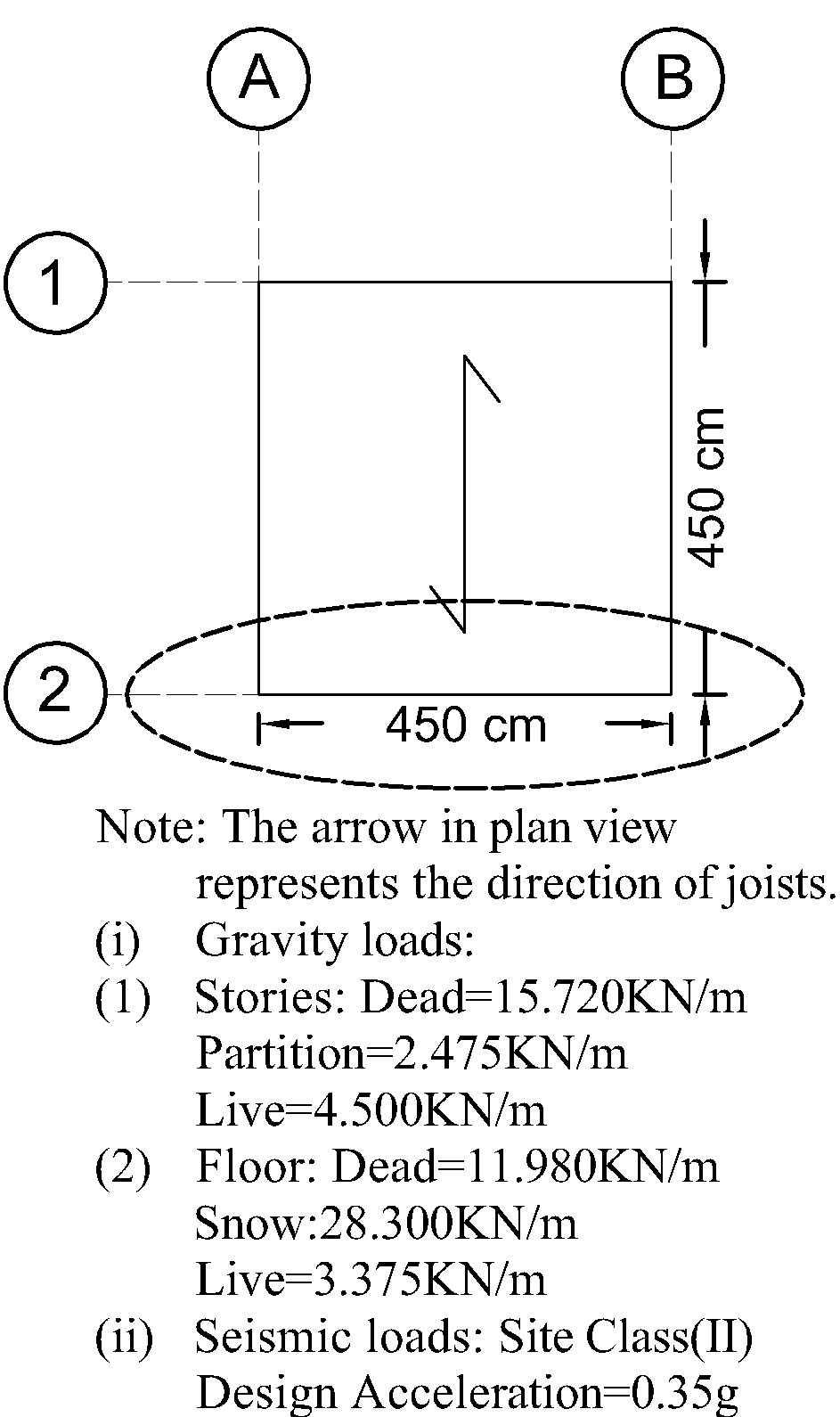}
         \caption{Plan schematic view of the one-bay frame}
                  \label{fig:plan}
     \end{subfigure}
       \begin{subfigure}[b]{0.40\textwidth}
         \centering
         \includegraphics[width=\textwidth]{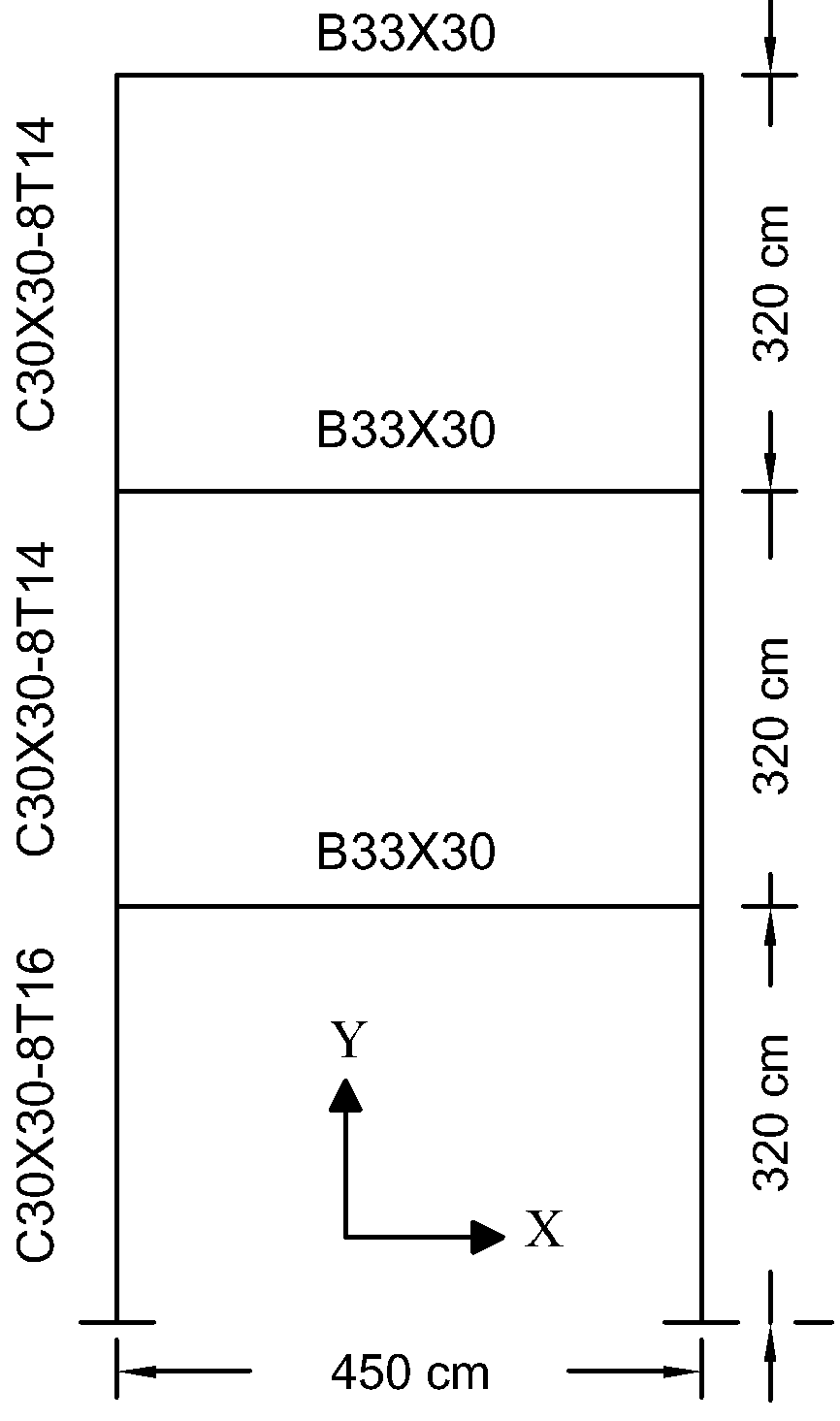}
         \caption{Front schematic view of the one-bay frame}
                 \label{fig:section}
     \end{subfigure}
\caption{(a) Plan schematic view of the one-bay frame (denoted within dashed ellipse) \\
and (b) front schematic view of the one-bay frame with design details.}
        \label{fig:foundation}
\end{figure}

On the basis of the prototype design, the RC frame is scaled to 1/2.78 of the actual structural size to meet the size limitations of the shaking table. The scaled model is designed and fabricated under the similitude law. The same materials are used in the prototype and scaled model structure. Additional masses (i.g., steel blocks) are affixed to the beams of each floor to represent the effective seismic weight and gravity loads acting on the floors. In order to meet similitude requirements, the total applied additional masses on the 1st - 3rd stories are 1320 kg, 1320 kg, and 1203 kg, respectively. Table 1 shows the similitude scale factors of the model parameters.
Details of the geometry of the model frame and the reinforcement are shown in Fig. 2. 
\begin{figure}[h]
     \centering
     \begin{subfigure}[b]{0.55\textwidth}
         \centering
         \includegraphics[width=\textwidth]{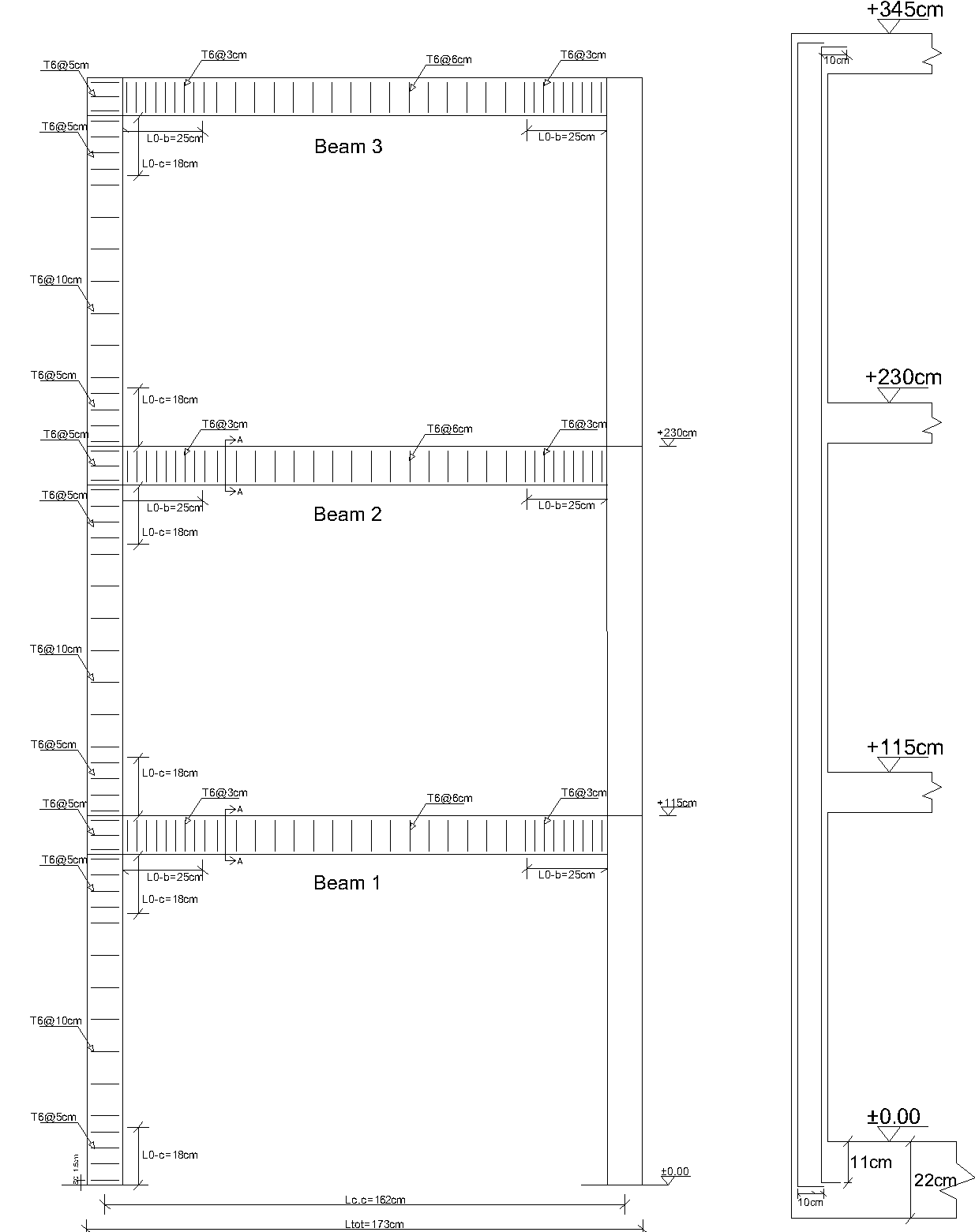}
         \caption{Elevation view of the frame and columns}
                  \label{fig:elev}
     \end{subfigure}
       \begin{subfigure}[b]{0.25\textwidth}
         \centering
         \includegraphics[width=\textwidth]{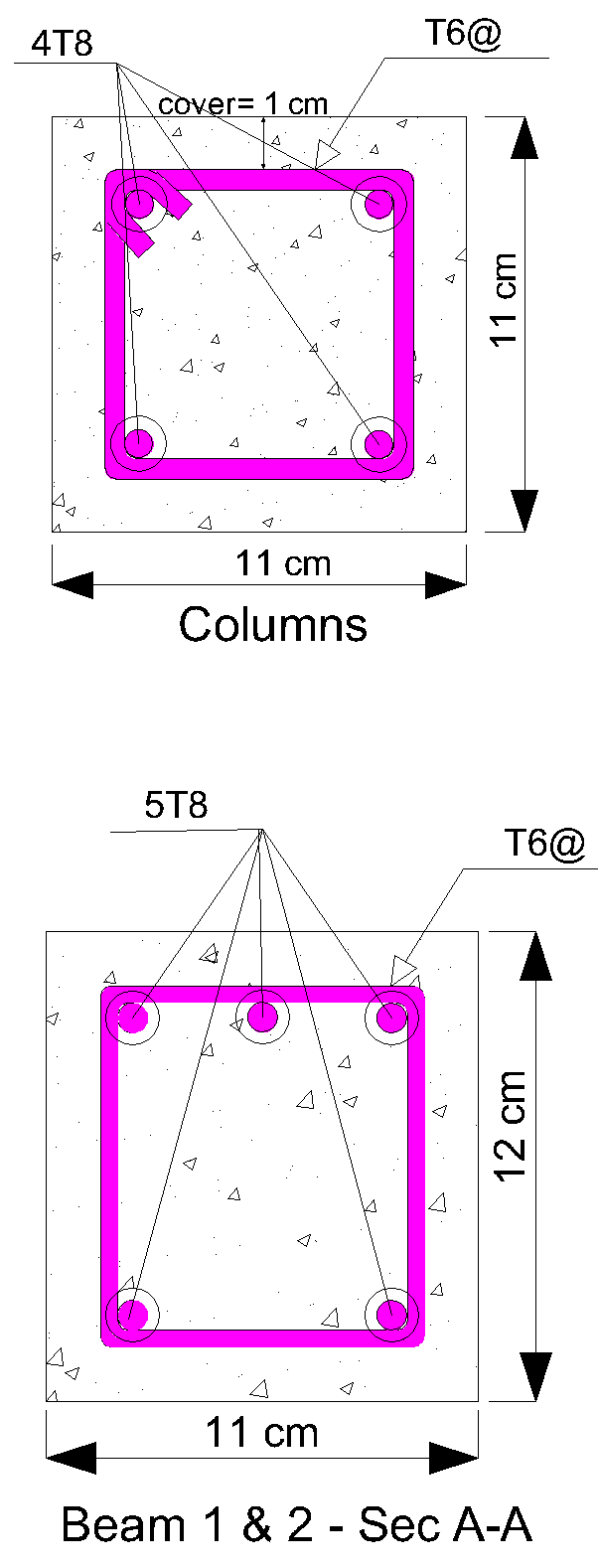}
         \caption{Members' cross-sections}
                 \label{fig:sections}
     \end{subfigure}
\caption{Specimen elevation and reinforcement detail of sections.}
        \label{fig: Model detail}
\end{figure}
The shaking table tests are conducted at the University of Tehran. The characteristics of the shaking table are summarized as follows: size (1.2×1.8 m); single degree of freedom; maximum stroke of the actuator (±125 mm); capacity of vertical load (50KN); frequency of input motion (0.01Hz - 18Hz); maximum base acceleration (1.0g).
\begin{table}[h!]
\centering
\caption{Similitude scale factors}
\begin{tabular}{C{3cm} C{3cm} C{3cm}} 
 \hline
   Parameter & Relationship & Model/Prototype  \\ [0.5ex] 
 \hline
 Length & $S_l$ & 1/2.78  \\ 
 Young’s modulus & $S_E$ & 1.00  \\
 Stress & {$S_\sigma=S_E$} & 1.00  \\
 Strain & {$S_\sigma/S_E$}  & 1.00  \\
 Time & ($S_m/S_k   )^{0.5}$  & 0.60  \\ 
 Frequency & $1/S_t $ & 1.67  \\
 Velocity & {$S_l/S_t$}  & 0.60  \\
 Acceleration & {$S_a$}  & 1.00  \\[1ex]
 \hline
 \end{tabular}
\label{table:1}
\end{table}

\subsection{Pinhole Camera Model}
 The pinhole camera model~\cite{ma2012imagemodel} is used to perform camera calibration and object reconstruction~\cite{sun2020measuring} from images. A point $\mathcal{P}$ with coordinates is represented as $\bm{X}_{P}=[X_{P},Y_{P},Z_{P}]^T$ in world coordinate system with Euclidean format.
A point in world coordinates system can be denoted using the Euclidean representation as $[\tensor*[^w]{X}{}, \tensor*[^w]{Y}{}, \tensor*[^w]{Z}{}]^{T}$. It can be projected onto the image plane defined by the sensor coordinate system $[\tensor*[^s]{x}{}, \tensor*[^s]{y}{}]^{T}$ using a perspective transformation:

\begin{equation}
	s\bm{m'}=\bm{A \left[ R|t \right] M} 
	\label{eq:cam_matrix} 
\end{equation}

\noindent where $\bm{A}$ is a matrix of intrinsic parameters of the camera, $\bm{[ R|t]}$ is the joint rotation-translation matrix (including extrinsic parameters), $s$ is scale factor for images, and $\bm{m'}=[\tensor*[^s]{x}{}, \tensor*[^s]{y}{}, {1}]^{T}$  and $\bm{M}=[\tensor*[^w]{X}{}, \tensor*[^w]{Y}{}, \tensor*[^w]{Z}{}, {1}]^{T}$ are the homogeneous coordinates in the sensor coordinate system and the world coordinate system, respectively.

The intrinsic matrix $\bm{A}$ and distortion coefficients (i.e. radial and tangential distortion) can be computed by using a chessboard calibration.  The extrinsic parameters of a camera can be computed by the Levenberg--Marquardt algorithm~\cite{more1978levenberg} relating the locations of objects in both world and sensor coordinates. In the study, the intrinsic parameters of the cameras are constructed using the checkerboard calibration and the extrinsic parameters are obtained from direct measurements.

\section{Experiment}
%
The seismic loading protocol for the study consists of 9 gradually increasing ground motion records. Sarpol-E-Zahab earthquake record obtained from the 2017 Sarpol-E-Zahab earthquake is used as the input excitation for the shake table test. The model frame is subjected to a sequence of gradually increasing excitations ranging from 0.1g to 0.6g along the X-axis as listed in Table 2. The main ground motion is time-scaled by the scale factor of $1/\sqrt{2.78}$ considering the model scaling factor of $1/2.78$. It is also amplitude-scaled to different amplitudes for different test stages. The time history of the acceleration and the spectrum of the ground motion are presented in Fig.3(a) and Fig.3(b), respectively.\\
  
  \begin{figure}
     \centering
     \begin{subfigure}[b]{0.48\textwidth}
         \centering
         \includegraphics[width=\textwidth]{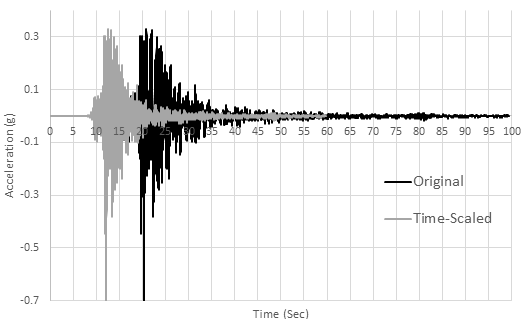}
         \caption{Acceleration time history}
         \label{fig:Acc}
     \end{subfigure}
     \begin{subfigure}[b]{0.48\textwidth}
         \centering
         \includegraphics[width=\textwidth]{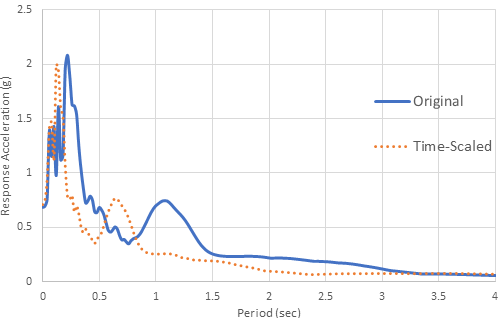}
         \caption{Acceleration response spectrum}
         \label{fig:Reinforcement}
     \end{subfigure}
        \caption{Time history of input acceleration and response spectrum of input ground motion.}
        \label{fig:Acc and response}
\end{figure}

\begin{table}[h!]
\centering
\caption{Details of input ground motions}
\begin{tabular}{C{1cm} C{3cm} C{1.5cm} C{2cm} } 
 \hline
  Case &  Name & Direction & Intensity(g)  \\ [0.5ex] 
 \hline
 1 & Sarpol-E-Zahab & X & 0.10  \\ 
 2 & Sarpol-E-Zahab & X & 0.20 \\ 
 3 & Sarpol-E-Zahab & X & 0.30 \\
 4 & Sarpol-E-Zahab & X & 0.35 \\
 5 & Sarpol-E-Zahab  & X & 0.40 \\
 6 & Sarpol-E-Zahab  & X & 0.45 \\ 
 7 & Sarpol-E-Zahab & X & 0.50 \\
 8 & Sarpol-E-Zahab  & X & 0.55 \\
 9 & Sarpol-E-Zahab  & X & 0.60 \\[1ex]
 \hline
\end{tabular}
\label{table:2}
\end{table}
The structural responses of the model structure that are monitored during the shake table test are the accelerations and displacements of stories. Three types of sensors are used in the test (i.e., accelerometer, LVDT, and strain gauges). Four accelerometers, among which three of them are installed on the beam of each floor and one is adopted on the shaking table. Four linear variable differential transformers (LVDTs) are used to acquire the displacement time history and drift of stories. Nine strain gauges are used to measure the strains of longitudinal bars. Strain gauges are attached on the surface of longitudinal bars at column and beam ends. In addition, three digital cameras record the experiment during the dynamic tests. 
Tension tests are conducted on samples of reinforcing bars, giving approximate yield stress of 404 MPa for the longitudinal reinforcement. Compression tests on the concrete samples resulted in 26.5 MPa for the 28-day concrete compressive strength.
The specimen is placed on the center of the shake table as shown in Fig.4(a) and it is subjected to selected input ground motions. In order to simulate the gravity and effective seismic loads on the frame and also to satisfy similitude requirements between the prototype and test model, steel blocks are attached to the beam of each story.

\begin{figure}
     \centering
     \begin{subfigure}[b]{0.40\textwidth}
         \centering
         \includegraphics[width=\textwidth]{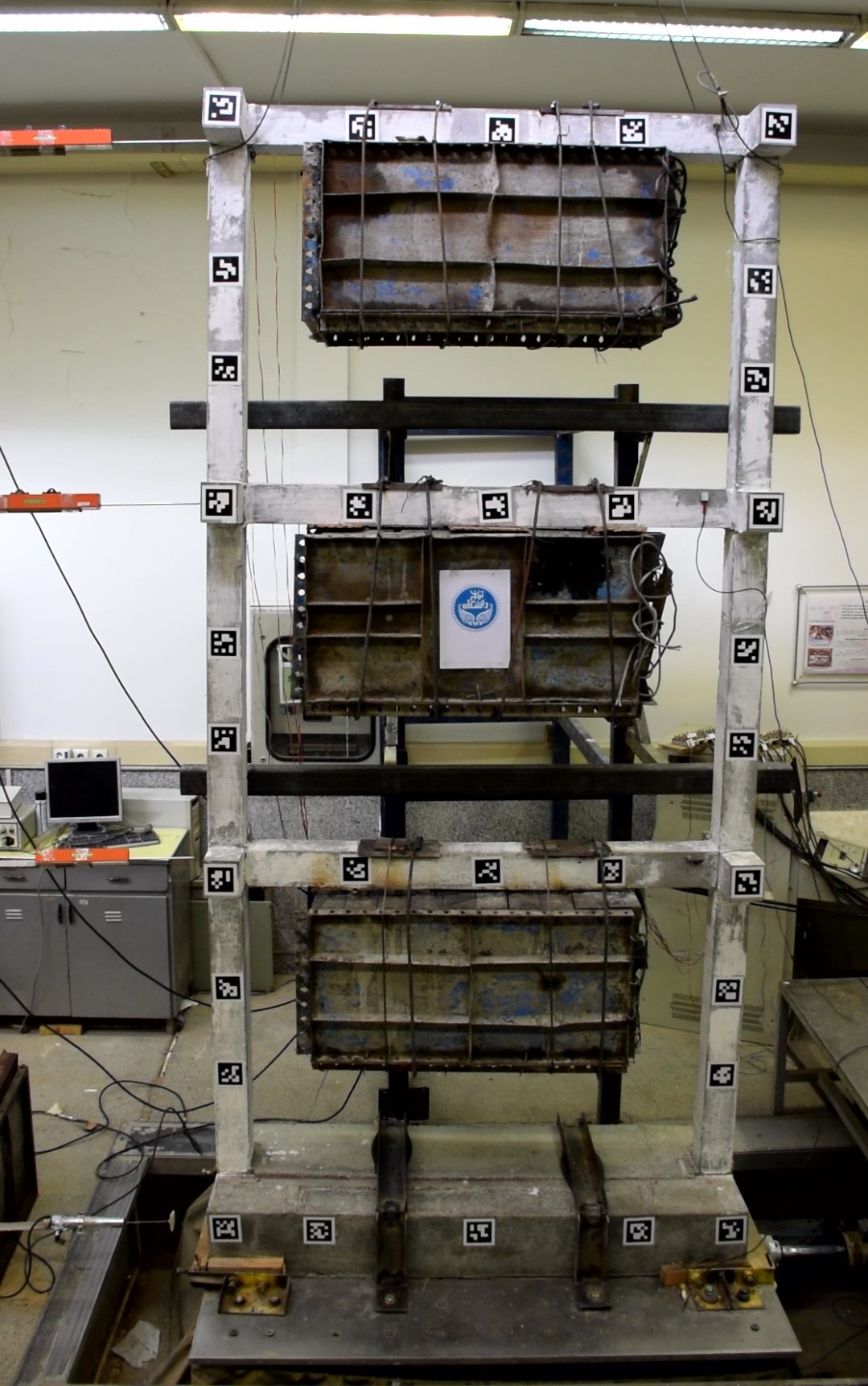}
         \caption{IMF model attached with AprilTags}
         \label{fig:test frame}
     \end{subfigure}
    \hfill
     \begin{subfigure}[b]{0.44\textwidth}
         \centering
         \includegraphics[width=\textwidth]{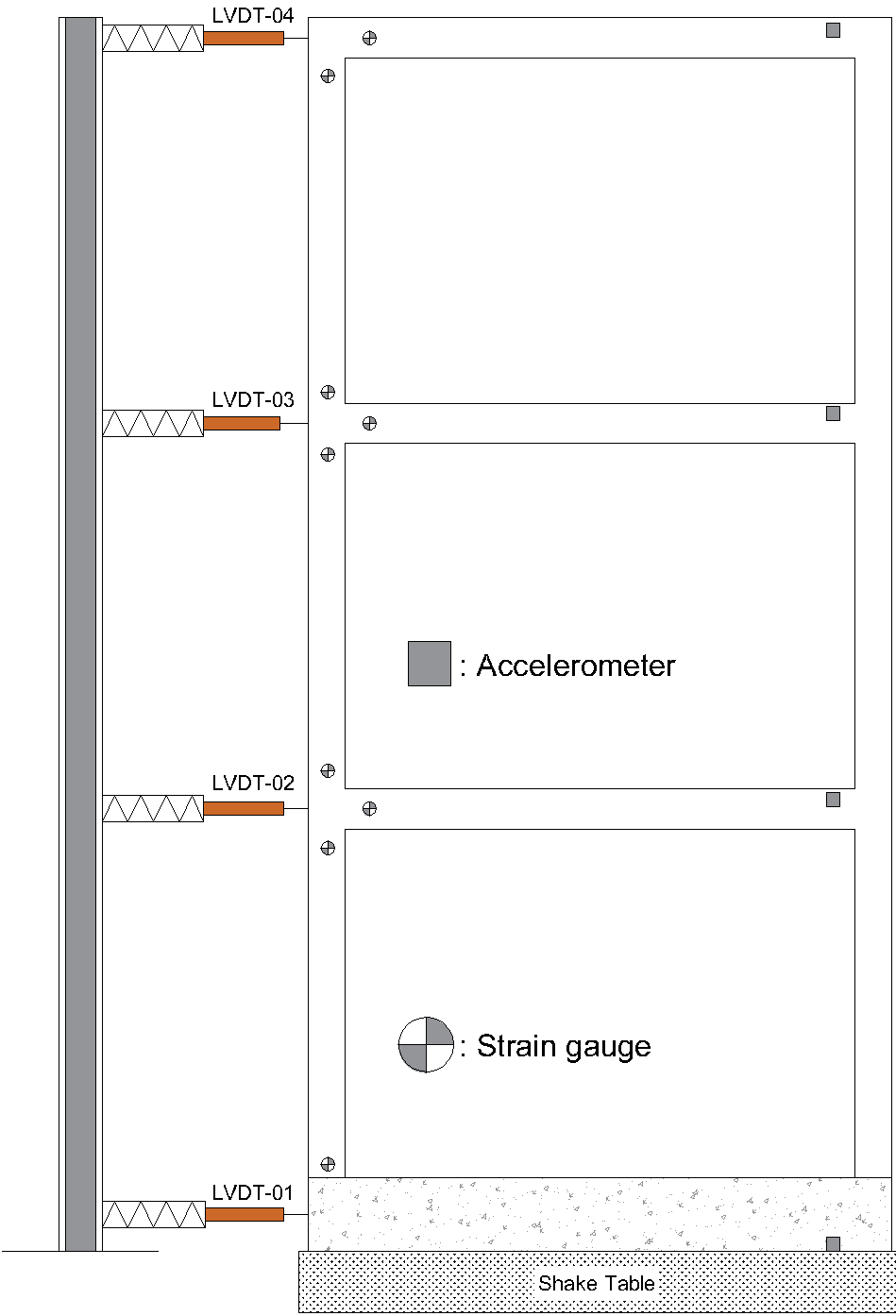}
         \caption{Deployment of conventional sensors}
         \label{fig:sensors}
     \end{subfigure}
       \begin{subfigure}[b]{0.8\textwidth}
         \centering
         \includegraphics[width=\textwidth]{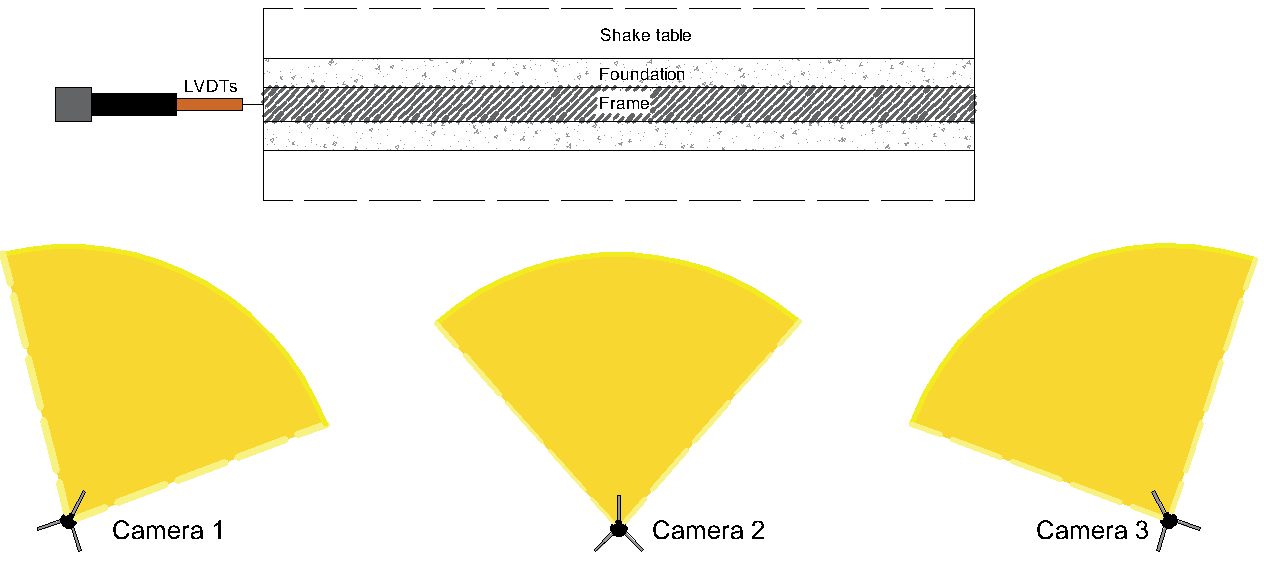}
         \caption{Plan view of cameras and frame}
         \label{fig:test frame plan}
     \end{subfigure}
        \caption{Experimental setup for the shaking table test.}
        \label{fig:test setup}
\end{figure}

%
\section{Future Work}
%
A Series of shaking table tests are implemented on the intermediate RC IMF. For each story, the seismic behavior (\textit{e.g.,} displacement, drift, and corresponding acceleration) will be discussed in future work. Moreover, the resulting dynamic characteristics will be compared between using data from conventional sensors and using data from vision-based sensors.

%

%
%

\bibliographystyle{splncs03_unsrt}
\bibliography{ShakeTable_v1.0.bib}

\end{document}